\documentstyle[emulateapj,apjfonts,psfig]{article}

\lefthead{J.~M.~Miller et al.}  
\righthead{A Hard Look at GX 339$-$4}

\received{Date}
\revised{Date}
\accepted{Date}

\journalid{vol}{date}
\articleid{1}{4}
\paperid{id}

\cpright{AAS}{1999}
\ccc{x}

\begin{document}

\title{A Long, Hard Look at the Low--Hard State in Accreting Black Holes}

\author{J.~M.~Miller\altaffilmark{1},
        J.~Homan\altaffilmark{2},
	D.~Steeghs\altaffilmark{3},
	M.~Rupen\altaffilmark{4},
	R.~W.~Hunstead\altaffilmark{5},
	R.~Wijnands\altaffilmark{6},
        P.~A.~Charles\altaffilmark{7}$^,$\altaffilmark{8},
        A.~C.~Fabian\altaffilmark{9}}

\altaffiltext{1}{Department of Astronomy, University of Michigan, 500
Church Street, Ann Arbor, MI 48109, jonmm@umich.edu}
\altaffiltext{2}{Kavli Institute for Astrophysics and Space Research,
MIT, 77 Massachusetts Avenue, Cambridge, MA 02139}
\altaffiltext{3}{Harvard-Smithsonian Center for Astrophysics, 60
	Garden Street, Cambridge, MA 02138}
\altaffiltext{4}{Array Operations Center, National Radio Astronomy
Observatory, 1003 Lopezville Road, Socorro, NM 87801}
\altaffiltext{5}{School of Physics A29, University of Sydney, NSW
2006, Australia}
\altaffiltext{6}{Astronomical Institute ``Anton Pannekoek'', Kruislaan
403, University of Amsterdam, 1098 SJ Amsterdam, NL}
\altaffiltext{7}{School of Physics \& Astronomy, University of
Southampton, Highfield Campus, Southampton SO17 1BJ, UK}
\altaffiltext{8}{South African Astronomical Observatory, PO Box 9,
Observatory 7935, Cape Town, South Africa}
\altaffiltext{9}{Institute of Astronomy, University of Cambridge,
Madingley Road, Cambridge CB3 OHA, UK}


\authoremail{jonmm@umich.edu}

\label{firstpage}

\begin{abstract}
We present the first results of coordinated multi-wavelength
observations of the Galactic black hole GX~339$-$4 in a canonical
low--hard state, obtained during its 2004 outburst.  {\it XMM-Newton}
observed the source for 2 revolutions, or approximately 280~ksec; {\it
RXTE} monitored the source throughout this long stare.  The resulting
data offer the best view yet obtained of the inner accretion flow
geometry in the low--hard state, which is thought to be analogous to
the geometry in low-luminosity active galactic nuclei.  The {\it
XMM-Newton} spectra clearly reveal the presence of a cool accretion
disk component, and a relativistic Fe~K emission line.  The results of
fits made to both components strongly suggest that a standard thin
disk remains at or near to the innermost stable circular orbit, at
least in bright phases of the low--hard state.  These findings indicate
that potential links between the inner disk radius and the onset
of a steady compact jet, and the paradigm of a radially--recessed
disk in the low--hard state, do not hold universally.  The results of
our observations can best be explained if a standard thin accretion
disk fuels a corona which is closely related to, or consistent with,
the base of a compact jet.  In a brief examination of archival data,
we show that Cygnus X-1 supports this picture of the low/hard
state.  We discuss our results within the context of disk--jet
connections and prevailing models for accretion onto black holes.
\end{abstract}

\section{Introduction}
The nature of accretion onto black holes at low fractions of the
Eddington mass accretion rate has been the topic of considerable
observational and theoretical attention.  Especially in the {\it
Chandra} and {\it XMM-Newton} era, improved sensitivity has permitted
the study of very faint sources.  {\it Chandra} observations of Sgr A*
have begun to probe the nature of the accretion flow onto this
super-massive black hole at $L_{X}/L_{Edd} \simeq 10^{-10}$ (Baganoff
et al.\ 2003).  Yet, the nature of accretion onto black holes even at
$L_{X}/L_{Edd} \simeq 10^{-2}$ remains uncertain.  Galactic black
hole transients are especially good laboratories in which to study the
manner in which flows change with mass accretion rate.  In particular,
clear changes in the mass accretion rate and flow geometry are thought
to be marked by changes in X-ray colors, flux, fast X-ray variability,
and multi-wavelength properties as a source transitions from a
high--soft state to the low--hard state (for reviews, see McClintock
\& Remillard 2005, Homan \& Belloni 2004).

Building upon analysis of spectral states by Miyamoto et al.\ (1995),
Homan et al. (2001) showed that simple variations in the implied mass
accretion rate alone cannot account for the state transitions and
properties observed in the 1998--1999 outburst of XTE~J1550$-$564.  It
was noted that at least one additional parameter appears to have a
profound influence on the flux and properties of an accreting black
hole.  Homan et al.\ (2001) suggested that the second parameter might
be related to the size of the corona.  The size of the corona might
naturally be regulated by the radius of the inner edge of an
optically-thick accretion disk.  Alternatively, the inner disk radius
may not change, or may be less important than independent
characteristics of the corona.

A common, paradigmatic model of how accretion flows change with
state and mass accretion rate was given by Esin et al.\ (1997).  In
that model, high--soft states are dominated by a standard thin
accretion disk, but in the low--hard state the inner disk is radially
truncated and replaced by an advection-dominated accretion flow.  The
same essential geometry is thought to describe quiescent phases.  The
fundamental assumption of a radially recessed accretion disk may find
some support in the low disk reflection fractions which are sometimes
measured in the low/hard state (e.g., Gierlinski et al.\ 1997).
Beloborodov (1999) proposed a very different picture for the nature of
state transitions and the accretion flow geometry in the low--hard
state.  In this model, black hole states are driven by the height of
magnetic flares above a disk which remains at the innermost stable
circular orbit (ISCO).  These flares would serve to feed a mildly
relativistically--outflowing corona.  Low disk reflection fractions do
not signal a recessed disk in this model, but result from mild beaming
of the hard X-ray flux away from the disk.  Merloni \& Fabian
(2002) have discussed a similar picture of the low/hard state, based
on magnetically--dominated coronae.

In order to better understand the nature of the low--hard state, we
proposed to make a deep observation of GX~339$-$4 in the low--hard
state with {\it XMM-Newton}.  GX~339$-$4 is a well-known recurrent
transient, which harbors a low-mass donor star and a black hole with a
mass of at least $5.8~M_{\odot}$ (Hynes et al.\ 2003).  The distance
to GX~339$-$4 is likely 8~kpc (Zdziarski et al.\ 2004; also see Hynes
et al.\ 2004).  It has been well-studied at all wavelengths, but
notable recent results include the detection of transient jets with
$v/c > 0.9$ (Gallo et al.\ 2004), evidence for a high black hole spin
parameter based on a prominent Fe~K$\alpha$ emission line in its X-ray
spectrum (Miller et al.\ 2004a, 2004b), and evidence for broad-band
jet emission based on multi-wavelength observations (Homan et al.\
2005).  In the sections that follow, we present the results of X-ray
spectral fits to GX~339$-$4 in its low--hard state.

\section{Observations and Data Reduction}
{\it XMM-Newton} observed GX~339$-$4 during revolutions 782 and 783,
for a total exposure of approximately 280~ksec.  Extensive
simultaneous radio observations were obtained with the Australia
Telescope Compact Array (ATCA).  Optical and IR monitoring
observations were obtained simultaneously at the South African
Astronomical Observatory (SAAO) using the 1.0m and IRSF 1.4m
telescopes.  In addition, the Optical Monitor (OM) onboard {\it
XMM-Newton} provided simultaneous V band observations at high time
resolution.  The full results of this multi-wavelength observing
program will be presented in a separate paper, and are only mentioned
briefly in this work.

The ``medium'' EPIC optical blocking filter was used during the
observations made with {\it XMM-Newton}, as the optical magnitude of
the source was expected to be fainter than $V=6$.  The prime instrument
for this observation was the EPIC-pn camera, which was run in timing
mode.  Unfortunately, the source flux was too high for this mode, and
the camera experienced photon pile-up.  We note that {\it XMM-Newton}
has since revised the flux and count rate limits for timing mode to be
a factor of two lower.  The RGS units were operated in the standard
``spectroscopy'' mode, and the OM was operated in ``fast'' mode.  Our
analysis of the data from these instruments will be presented in
separate work.

The EPIC-MOS1 and EPIC-MOS2 cameras were operated in the standard
``full frame'' mode.  The MOS cameras have a nominal operational range
of 0.2--10.0~keV.  Data from these cameras was reduced using SAS
version 6.5.  The EPIC-MOS camera good time started on 2004--03--16 at
16:23:41 (TT).  Although these cameras also suffered photon pile-up,
it was possible to extract robust spectra and lightcurves from these
cameras by using annular extraction regions to avoid the piled-up core
of the bright source image (spatial information is not preserved in
EPIC-pn ``timing'' mode).  The spectra obtained from these cameras are
the centerpiece of the analysis detailed in this work.  Using the SAS
tool ``xmmselect'' we filtered the MOS1 and MOS2 event lists to
include event grades 1--12 and ``flag$=$0''.  The event lists were
screened against times with high instrumental flaring.  This resulted
in net MOS1 and MOS2 exposure times of 81~ksec (each) during
revolution 782, and 59~ksec (each) during revolution 783.  Using the
xmmselect tool ``epatplot'', we compared the spectra extracted in a
number of annuli to the expected distribution of energy and event
grades with radius.  Via this procedure, it was found that extracting
source counts in annuli between 18--120'' resulted in spectra free
from pile-up.  Background spectra were taken from a 60'' circle near
the corner of the central chip on each MOS camera.  The ``backscale''
tool was used to correctly normalize the source and background events.
Finally, the FTOOL ``grppha'' was used to require at least 10 counts
per bin in each spectrum.  Redistribution matrix files (rmfs) and
ancillary response files (arfs) for each spectrum were calculated
using the tools ``rmfgen'' and ``arfgen''.

During the long {\it XMM-Newton} stare, simultaneous X-ray snapshots
were obtained with {\it RXTE}, in order to better-define the
broad-band X-ray continuum emission and to characterize the timing
properties of the source.  These observations were reduced using the
packages and tools available in HEASOFT version 6.0. As this paper is
focused on the results of fits made to the {\it XMM-Newton}/EPIC-MOS
spectra, we have chosen to focus on one representative {\it RXTE}
observation.  Observation 90118-01-06-00 began on 2004-03-17 at
12:03:12 (TT).  After standard screening (e.g., against SAA
intervals), net PCA and HEXTE exposures of 2.2~ksec and 0.8~ksec were
obtained.  

PCU-2 is the best-calibrated PCU in the {\it RXTE}/PCA.  A spectrum
from PCU-2 was extracted from data taken in ``Standard2'' mode,
providing full coverage of the 2.0-60.0~keV bandpass in 129 channels
every 16 seconds.  Data from all the Xe gas layers in PCU-2 were
combined.  Background spectra were made using the FTOOL ``pcabackest''
using the latest ``bright source'' background model.  An instrument
response file was obtained using the tool ``pcarsp''.  It is
well-known that fits to PCA spectra of the Crab with a simple
power-law model reveal residuals as large as 1\%, but adding 1\%
errors to PCA data is often an over-estimate (see, e.g., Rossi et al.\
2005).  Using the tool ``grppha'', we added 0.6\% systematic errors to
the PCU-2 spectra.  The HEXTE-A cluster was operated in the standard
``archive'' mode, which has a time resolution of 32~s and covers the
10.0--250.0~keV band with 61 channels.  We extracted a
background-subtracted spectrum from the HEXTE-A cluster, and an
associated instrument response file, using the standard procedures.

The {\it XMM-Newton}/EPIC-MOS cameras nominally cover the
0.2--10.0~keV band.  A preliminary inspection of the data revealed
small but significant deviations at the lowest and highest energy
bounds.  These likely result from small anomalies in the instrument
response function for the annuli we have used; we therefore restricted
our spectral analysis to the 0.7--9.0~keV band.  The low energy bins
in the {\it RXTE}/PCA are not calibrated as well as the central part
of the bandpass, and the same is true for the higher energy portion of
the bandpass.  Therefore, we restricted our analysis of the PCU-2
spectrum to the 2.8--25.0~keV band.  Prior experience with PCA spectra
suggests that the strongest Xe L edge is not fully accounted for by
the detector response; in all fits to the PCU-2 spectrum detailed
below, an edge at 4.78~keV with $\tau=0.1$ was included to account for
this small effect.  We restricted our analysis of the HEXTE-A spectrum
to energies above 20~keV.  The HEXTE-A spectrum has little signal
above 100~keV, so this energy was set as an upper bound on all fits.
XSPEC version 11.3.2 was used to analyze the {\it XMM-Newton} and {\it
RXTE} spectra.  All errors reported in this work are 90\% confidence
errors obtained by allowing all parameters to vary, unless otherwise noted.

\section{Analysis and Results}
\subsection{ATCA Radio Observations}

Throughout the long stare obtained with {\it XMM-Newton}, GX~339$-$4
was observed with the Australia Telescope Compact Array (ATCA) at 4.8
and 6.2~GHz, for the bulk of 14-hour observing runs between 16 and 19
March 2004.  The data were reduced in the standard fashion using the
Miriad (e.g., Sault \& Killeen 2004) and AIPS (e.g., Greisen 2005)
software packages.  The resulting images showed GX~339$-$4 to be a
point source with no obvious extension, at a resolution of $6.1\times
5.2$\arcsec (4.8 GHz), and $5.7\times 4.9$\arcsec (6.2 GHz).  The mean
flux densities (referenced to the standard flux calibrator PKS
B1934$-$638) were $5.3\pm 0.05~\rm mJy$ (4.8 GHz) and $5.7\pm 0.05~\rm
mJy$ (6.2 GHz), where the error bars reflect the rms noise in the
images from individual days, and there is no sign of strong
variability.

In summary, ratio of radio flux to X-ray flux, the radio spectral
index ($\alpha = +0.3\pm 0.1$, $S_\nu\propto\nu^\alpha$), and the lack
of strong variability (on the order of 0.1~mJy from day to day), are
all typical of radio emission in the low--hard state.  Such radio
emission is generally interpreted as the signature of a steady,
compact jet (e.g., Fender 2005).

\subsection{RXTE Timing Analysis}

An inspection of the {\it RXTE} X-ray light curves revealed strong
flaring, superimposed on a relatively stable base flux level.  The flares
occurred on a time scale of a few tens of seconds and the flare maxima
occasionally reached count rates close to eight times that of the base
level.  Power spectra were created from the GoodXenon data (total
energy band, 2--60 keV), with frequency ranges of
$128^{-1}$--1024 Hz.  The power spectra were normalized according to
Belloni and Hasinger (1990) and Miyamoto et al. (1991), and the
Poisson level was subtracted following the method described in
Klein-Wolt et al. (2004).  

Two broad peaks could be identified in the power spectrum of each of
the {\it RXTE} observations (a representative power spectrum is shown
in Figure 1).  One Lorentzian peaks around 0.05 Hz, corresponding to
the strong flaring that is directly visible in the light curves, and a
second, broader feature peaks around 2 Hz.  The low-frequency peak
power could be fit well with one Lorentzian, with a Q-value
(i.e. frequency divided by full-width-at-half-maximum) of $\sim$0.3
and a fractional rms amplitude of $\sim$30\%.  This feature can be
identified with the break that is seen in the hard state power spectra
of, e.g. Cyg X-1 (Belloni \& Hasinger 1990).  Two Lorentzians were
needed to fit the feature at higher frequencies, peaking around 0.5 Hz
(Q fixed to 0, rms$\approx$20\%) and 2.5 Hz (Q$\approx$0.15,
rms$\approx$25\%).  The total strength of the variability in the
$128^{-1}$--64 Hz range is about 47--51\%.  The variability proporties
in our observations are consistent with those of the canonical hard
state.

\subsection{Preliminary RXTE Spectral Analysis}

We made a preliminary study of the {\it RXTE} PCU-2 and HEXTE-A
spectra.  The spectra were fit jointly within XSPEC, utilizing a
constant to account for normalization differences.  The continuum is
well fit by a very simple model consisting of a power-law modified by
absorption (using the ``phabs'' model with $N_{H}$ fixed at $4.0\times
10^{21}~{\rm atoms}~{\rm cm}^{-2}$, Miller et al.\ 2004a).  Only a
broad Fe~K$\alpha$ emission line (see Figure 2) prevents the simple
power-law model from being a formally acceptable fit ($\chi^{2}/\nu =
385.7/76$).  Using this continuum model, we measure a photon index of
$\Gamma = 1.50(2)$, and a normalization of $0.21(1)~{\rm ph}~{\rm
cm}^{-2}~{\rm s}^{-1}~{\rm keV}^{-1}$ at 1~keV.  This corresponds to
an unabsorbed flux of $5.33\times 10^{-9}~{\rm erg}~{\rm cm}^{-2}~{\rm
s}^{-1}$ in the 3--100~keV band, or a luminosity of $4.1\times
10^{37}~{\rm erg}~{\rm s}^{-1}$ (or $L_{X} \leq 0.05~L_{Edd}$) for a
distance of 8~kpc.  While there is some evidence for subtle curvature
in the high energy spectrum that is consistent with disk reflection,
it is too weak to be significant.  Power-law models with an
exponential high-energy cut-off and broken power-law models do not
provide significant improvements to the continuum fit.

A simple power-law fit in th 20--100~keV band gives an energy flux of
$2.5\times 10^{-9}~{\rm erg}~{\rm cm}^{-2}~{\rm s}^{-1}$, and a photon
flux of $3.6\times 10^{-2}~{\rm ph}~{\rm cm}^{-2}~{\rm s}^{-1}$.  This
places our simultaneous X-ray and radio flux measurements in the
middle of the radio--X-ray flux correlation Corbel et al.\ (2000)
found to hold in the low--hard state of GX~339$-$4, and further
indicates that we observed GX~339$-$4 in a standard low--hard state.


\subsection{Preliminary XMM-Newton Spectral Analysis}

We next explored simple continuum fits to the {\it
XMM-Newton}/EPIC-MOS spectra.  We fit the four spectra (MOS1 and MOS2
from revolutions 782 and 783) jointly, allowing a normalizing constant
to float between the spectra.  A simple power-law model like that
which adequately described the {\it RXTE} spectra was unable to fit
the continuum in the {\it XMM-Newton} spectra, due to a strong soft
flux excess.  

The presence of a strong soft excess in the {\it XMM-Newton} spectra
does not represent a flaw in the performance or calibration of the
instruments aboard either observatory.  Rather, the soft flux excess
is merely not required in the {\it RXTE} band, due to its effective
low-energy bound of 3~keV.  When a low energy threshold of 3.0~keV is
chosen to match that of {\it RXTE}, a simple power-law model
adequately describes the continuum in the {\it XMM-Newton} spectra.
Extending this fit down to 0.7 keV reveals that the soft excess is
very significant below 2--3~keV in the {\it XMM-Newton} spectra.  This
effect is illustrated in Figure 3.

For any reasonable continuum model (a continuum consisting of one or
two additive components, which can be strongly constrained by the
data), fits reveal strong evidence for a relativistic iron line
arising from the inner disk (see Figures 4, 5, and 6). 

Studies of black hole X-ray binary outbursts with {\it RXTE} have
documented falling accretion disk flux and apparent temperature
through the outburst decay (e.g., Park et al.\ 2004; see also
McClintock \& Remillard 2005).  Below a certain disk temperature and
flux level, however, {\it RXTE} is simply unable to detect disk
emission.  In plotting additive component fluxes measured across an
outburst with {\it RXTE}, then, it is not uncommon to see a sudden
disappearance of the disk flux.  However, it is unlikely that disk
emission simply turns off at this point.  It is also unlikely that
{\it RXTE} can provide robust constraints on the nature of the disk in
low-temperature, low-flux phases.  It is most likely that the soft
excess we have discovered with {\it XMM-Newton} is due to emission
from an optically-thin, geometrically-thick accretion disk, and that
{\it XMM-Newton} is much better-suited to such measurements.

\subsection{Joint Spectral Fits with Simple Models}

We explored a number of fits with different disk components and hard
components, to demonstrate that the disk and disk line components
truly arise from an accretion disk, and are not modeling artifacts.
We note that no fits to the {\it XMM-Newton}/EPIC-MOS spectra, by
themselves or in combination with {\it RXTE} spectra, are formally
acceptable.  This is due to the presence of residual instrumental
response deficiencies near 1~keV and 2~keV.  Similar response issues
have been documented previously (e.g., Miller et al.\ 2004b), and do
not complicate efforts to obtain strong constraints on the nature of
the accretion flow.  The results of this analysis are particularly
robust because the fits span the 0.7--100.0~keV range, providing
excellent constraints on the continuum emission, and because of the
extraordinary depth of our observations.

For each model fit to the spectra, a normalizing constant was included
to allow for flux differences between {\it XMM-Newton} and {\it RXTE}.
This constant was fixed at unity for each of the MOS cameras, but
component normalizations (not, e.g., temperature or photon index) were
allowed to float between the cameras to account for more minor flux
variations (2--5\% or less).  As the PCA and HEXTE spectra are not
suited to constraining the column density or the parameters of any
cool disk components, all soft component parameters were tied to the
values obtained for the MOS-1 spectrum from revolution 782.  This
procedure is reasonable as global disk parameters should only vary on
the viscous timescale (see, however, Belloni et al.\ 1997 and Vadawale
et al.\ 2003), which is longer than the timescale of our observations.
Apart from their normalizations, all emission line and hard component
parameters in fits to the PCA and HEXTE spectra were linked with
parameters measured via the MOS spectra.  The same normalizations were
used for both the PCA and HEXTE spectra, again by tying these
parameters together.  

The parameters measured with four plausible continuum models are
listed in Table 1 (see also Figures 4, 5, and 6).  The most important
results obtained from these fits are that a cool $kT \simeq 0.3$~keV
disk (cool compared to the $kT \simeq$1--2~keV disks observed in high
flux states; see McClintock \& Remillard 2005) and a strong
relativistic iron line are required in each model.  Using the F-test,
we find that both components are individually required at much more
than the 8$\sigma$ level of confidence.  Adding a smeared edge
component (``smedge'' in XSPEC; although unphysical, it is sometimes
included to approximate a disk reflection continuum) did not
significantly improve the fits.  This may indicate that the disk
reflection is very weak (see below).

In the three continuum models wherein the disk radius may be inferred
from the additive disk component, each fit points to an inner disk
extending close to or within $6~GM/c^{2}$.  We made fits with two
different disk models to ensure that any such finding is not strongly
dependent on the model chosen.  Inner disk radii inferred via disk
continuum fits are notoriously unreliable, as they depend on the hard
component assumed, the inner disk boundary condition, the column
density, spectral hardening, and other effects (see Zimmerman et al.\
2004 for a discussion of different disk models; see Merloni, Fabian,
\& Ross 2000 for a detailed discussion of hardening effects).  In all
four models, however, the ``Laor'' relativistic line model points to a
disk which may extend to $3~GM/c^{2}$.  This finding may provide
additional evidence that GX~339$-$4 harbors a spinning black hole
(Miller et al.\ 2004a, 2004b).

Three of the models considered include additive model components; the
fourth model (the ``bulk motion Comptonization'' or BMC model; Shrader
\& Titarchuk 1999) is a single continuum which attempts to account for
up-scattering from a disk self-consistently.  The fact that a
relativistic disk line component is still strongly required in fits
with the bulk motion Comptonization (BMC) model, and that the line
parameters are consistent with those measured via the additive models,
signals that the line itself is not an artifact of using additive
components to fit the X-ray continuum.  Moreover, the measured values
of the line are similar in each model.  We note that the disk
temperature derived via the BMC model ($kT = 0.26(1)$~keV) is at most
0.1~keV lower than values obtained via the fits with additive
components, and that the best fit is obtained with the canonical
``diskbb'' plus ``power-law'' continuum.  This finding may indicate
that the disk emission is not strongly Comptonized, and that any
coronal geometry may not strongly distort our view of the inner disk.

Simple theoretical considerations strongly support interpreting the
cool soft excess we have observed as an accretion disk.  For standard
thin accrection disks, $T \propto \dot{M}^{1/4}$ (assuming that $L
\propto \dot{M}$; e.g., Frank, King, \& Raine 2002).  Based on an {\it
XMM-Newton} observation of GX~339$-$4 in the ``very high'' or
``steep power-law state'', Miller et al.\ (2004b) report an apparent
disk temperature of $kT = 0.76$~keV for an unabsorbed disk flux of
$1.4\times 10^{-8}~{\rm erg}~{\rm cm}^{2}~{\rm s}^{-1}$
(0.5--10.0~keV).  In the low--hard state observation described in this
work, the unabsorbed disk flux is 10\% of the total, or $2.7\times
10^{-10}~{\rm erg}~{\rm cm}^{2}~{\rm s}$, and the range of disk
temperatures is $kT =$ 0.27--0.38~keV.  The ratio of apparent
temperatures and mass accretion rates is fully consistent with the $T
\propto \dot{M}^{1/4}$ relation expected for standard thin accretion
disks which extend to the innermost stable circular orbit.

While the results of these fits with simple models clearly demonstrate
the presence and robustness of a broad Fe~K emission line, it is also
worth noting that this line profile clearly reveals the
importance of dynamical broadening.  The line profile (see Fig.\ 5 and
Fig.\ 6) is similar to the classic, two-horn profile originally found in
the spectrum of MCG--6-30-15 with {\it ASCA} (Tanaka et al.\ 1995).
In one sense, this shape more clearly reveals that the line is
broadened by dynamical processes, than extremely skewed lines: at
least in the case of Seyfert AGN, the extreme red wing in some lines
may be partially affected by (though not due to) low-energy absorption
(e.g., Vaughan \& Fabian 2004).  Similarly, this line shape, and
the absence of Fe XXV and Fe XXVI absorption lines in the spectrum,
serve to rule-out high energy absorption as a contribution to the line
shape and flux (e.g., Done \& Gierlinski 2006).  Prior work has
discussed alternative broadening mechanisms extensively.  It has been
shown that Comptonization is extremely unlikely to cause the
broadening observed (e.g., Fabian et al.\ 1995, Reynolds \& Wilms
2000).  Velocity shifts and/or scattering effects in a jet or outflow
(e.g., Titarchuk, Kazanas, \& Becker 2003) have also been shown to be
extremely unlikely sources of line broadening (Fabian et
al.\ 1995; Miller et al.\ 2004a).

The fits we have made with relativistic line models imply a low inner
disk inclination in GX~339$-$4.  An inclination of $i \simeq
18^{\circ}$ is consistent with the fits obtained, and prior fits to
the line profile (Miller et al.\ 2004a,b).  If the binary is viewed at
this low inclination, an implausibly high black hole mass would be
implied (Hynes et al.\ 2003).  However, the inner disk and binary
plane can be misaligned in black hole binaries.  In at least two
systems (GRO~J1655$-$40 and V4641~Sgr), the inner disk inclination
implied by relativistic jets and the orbital plane of the binary are
known to be significantly different (see, e.g., Maccarone 2002).  It
is likely that GX~339$-$4 is another such system.  A velocity of
0.9$c$ is sufficient to create the Doppler boosting needed to make the
approaching jet visible and the receding jet invisible within an angle
of $26^{\circ}$, in radio observations of jets launched from
GX~339$-$4 (Gallo et al.\ 2004).  This implies that the inner disk is
viewed at an inclination of $26^{\circ}$ or less, since the jet velocity
may be higher and the beaming angle tighter (Gallo et al.\ 2004).  The
inclination implied by our spectral fits is broadly consistent with
that inferred from jets.  If the inclination is instead fixed at a
value as high as $55^{\circ}$, the inner radius implied by fits with
Laor line components is still within $6~r_{g}$, although this
inclination forces a significantly worse fit statistically.

\subsection{Joint Spectral Fits with Disk Reflection Models}

As the results of spectral fits with simple models clearly reveal an
accretion disk and relativistic disk line, we next attempted fits with
disk reflection models.  The methodology outlined above was
employed, with regard to multiplicative constants and normalizations.
Reflection models predict that an Fe~K emission line is merely the
most prominent part of the interaction between hard X-rays and the
accretion disk; more subtle effects include a small flux decrement
above the line energy, and a ``reflection hump'' which normally peaks
between 20--30~keV.  Fitting disk reflection models, then, represents
the most self-consistent means of treating the relativistic line and
hard X-ray continuum.

We considered two disk reflection models: the ``constant density
ionized disk'' model with solar abundances (CDID; Ballantyne, Iwasawa,
\& Fabian 2001), and ``pexriv'' (Magdziarz \& Zdziarski 1995).  It is
expected that the CDID model is more physical, in that it is
suited to high disk ionization and includes the effects of
Comptonization on absorption edges.  We therefore used the results of
fits with the CDID model as a guide when fitting ``pexriv'', which has
more free parameters.  In particular, we fixed the disk temperature
and ionization parameter within ``pexriv'' to values consistent with
those obtained when fitting the CDID model.  It should also be noted
that whereas the CDID model explicitly includes Fe line emission,
pexriv does not, and an additional Laor line component was included in
fits with pexriv.  Finally, reflection models are calculated in the
fluid frame, and to match the observed spectrum, a reflection model
must be convolved with the relativistic effects expected in the inner
disk around a black hole.  These effects expected around a spinning
black hole are described by the ``laor'' line model, and so we
convolved the CDID and pexriv models with the laor line element in all
spectral fits.  This procedure is referred to as relativistic
blurring.  In the case of fits with pexriv, the blurring parameters
were simply tied to the independent line component parameters.  

Fitting blurred reflection models is computationally intensive; the
fits themselves and error scans can take several days to calculate
even on high--end  workstations.  We therefore took the
pragmatic step of only including the {\it XMM-Newton}/EPIC-MOS spectra
from revolution 782.  The spectra from revolution 782 have more counts
than those from 783, as there was more observing time free from
instrumental flaring.

It is useful to compare our current results to those obtained from
reflection fits to bright hard states.  The CDID and pexriv reflection
models were fit to the {\it XMM-Newton}/EPIC-pn spectra of
XTE~J1650$-$500 and GX~339$-$4 in the ``very high'' or ``steep
power-law'' state.  In those cases, reflection fractions ($f \propto
\Omega/2\pi$) broadly consistent with unity and high disk ionizations
(log($\xi$) $\simeq$ 4.0, and higher) were measured (Miller et al.\
2002a; Miller et al.\ 2004b).  These prior results imply that in each
case, an ionized thin accretion disk extended to the innermost stable
circular orbit (ISCO).

The complete set of parameters obtained with the blurred constant
density ionized disk and pexriv models are detailed in Table 2, and
shown in Fig.\ 7 and Fig.\ 8.  Fits to the low--hard state spectra
from GX~339$-$4 suggest a moderate disk ionization, log($\xi$)
$\simeq$3.0 ($\xi = L_{X}/nr^{2}$, where $n$ is the hydrogen number
density).  This is consistent with the low apparent disk temperature
we have measured, and with the hard ionizing flux levels that are
observed in the low--hard state.  The low reflection fractions we have
measured ($f \simeq 0.2-0.3$) are consistent with prior results
obtained in the low--hard state (e.g., Gierlinski et al.\ 1997).

Whereas low disk reflection fractions could previously be explained in
terms of a radially-recessed accretion disk, our data appear to rule
out such an explanation.  The apparently inconsistent findings of a
disk remaining at the innermost stable circular orbit, and a low
reflection fraction, can be reconciled if the hard X-ray emission is
mildly beamed away from the accretion disk.  As with the case of
phenomenological models discussed above and detailed in Table 1, we
found that the parameter values obtained with reflection models are
not strongly dependent on the assumed inner disk inclination.

From the soruce luminosity and the ionization parameter measured via
the CDID model, it is possible to estimate the density of the
reflecting medium.  Assuming an inner radius of $3.0~r_{g}$, a density
of $n \sim 10^{21}~{\rm cm}^{-3}$ is implied.  Further assuming a
radiative efficiency of 0.1, continuity in the accretion flow means
that the product of the radial inflow velocity $v_{\rm r}$ and the
relative thickness of the disk $h/r$ is about $10^{7}$.  Such a small
value for this product means that we are dealing with a dense thin
disk ($h/r$ small) where the radial inflow velocity is much smaller
than the tangential, Keplerian one, $v_{\rm r} \ll v_{\phi} \sim c$.
For an $\alpha$-disk, $v_{\rm r}\sim \alpha v_{\phi} (h/r)^{2}$, so
$\alpha v_{\phi} (h/r)^{3} \sim 3 \times 10^{-4}$.  This is well
satisfied if $\alpha\sim 0.3$ and $(h/r)\sim 0.1$ as expected for a
standard thin disk.  Of course, disk reflection only deals with the
density near the surface of the disk; a higher mean disk density only
strengthens the result.  Thus, our reflection fits also strongly
suggest that there is something approximating a standard thin disk at
small radii.

We note that our results do not place strong constraints on the
dominant hard X-ray emission mechanism(s).  It is possible that
thermal Comptonization, synchrotron emission, and synchrotron
self-Comptonization all play important roles in defining the hard
X-ray emission in the low--hard state.  The apparent lack of a clear
break or cut-off in the high energy spectrum may argue that thermal
Comptonization is not the dominant emission mechanism, or may simply
argue for a coronal electron temperature in excess of 100~keV.

\subsection{Comparison with Cygnus X-1}

In view of the remarkable results detailed above, we undertook a brief
investigation of archival low--hard state Galactic black hole spectra.
For this purpose, we selected the well-known source Cygnus X-1. 
While Cygnus X-1 caused photon pile-up in the {\it ASCA}/SIS
detectors, robust spectra were obtained with the GIS gas detectors,
which were well-calibrated over the 1.0--10.0~keV bandpass.  {\it
ASCA}/GIS spectra and response files available in the NASA public
archive have been screened in accordance with the bright source
considerations detailed in Brandt et al.\ (1996), and are ready for
analysis.  Observation 40027010 was obtained on 11 November 1993
starting at 01:37:04; the GIS-3 spectrum considered here has a net
exposure time of 38~ksec.  

Assuming a standard value for the column density along the line of
sight to Cygnus X-1 ($N_{H} = 6.2\times 10^{21}~{\rm cm}^{-2}$; Miller
et al.\ 2002b), a power-law fit to the spectrum above 3.0~keV yields a
$\Gamma = 1.77$ photon power-law index typical of the low--hard state
in Cygnus X-1.  Extending this power-law down to 1.0~keV reveals a
strong cool accretion disk component like that found in GX~339$-$4
(see Figure 9).  Moreover, a broad, relativistic Fe~K$\alpha$ emission
line is revealed, which is again suggestive of a disk which extends to
$6~GM/c^{2}$ or less (see Figure 10).

Before including a disk component, the best fit statistic that can be
achieved with this power-law index is $\chi^{2}/\nu > 15,000/764$.
Allowing the column density to float within 30\% of the standard
value, which is a common deviation between instruments, an improved
fit was again found.  Next including ``diskbb'' and ``laor'' line
components each separately improved the fit at more than the 8$\sigma$
level confidence.  This simple model yielded the following parameters:
$kT = 0.22(1)$~keV for $R_{in} = 5.8(6)~R_{g}$ (assuming $d=2.5$~kpc,
$i=30^{\circ}$, and $M=10~M_{\odot}$, Miller et al.\ 2002b, Herrero et al.\
1995), $E_{line} = 6.90(7)$~keV for $R_{line} = 6(2)~R_{g}$, $W_{line}
= 250(30)$~eV, $\Gamma = 1.85(5)$, and $\chi^{2}/\nu = 1317/759$.
Again, radii inferred from disk continuum fits must be regarded
cautiously, though in this case the radius is still broadly consistent
with the ISCO after inner torque and hardening corrections are made
(see Zimmerman et al.\ 2004; Merloni, Fabian, \& Ross 2000).  The fit
is not formally acceptable due to instrumental response errors around
2~keV; however, the statistical need for disk and disk line components
does not arise due to response issues.

To explore the low/hard state at a lower fraction of the Eddington
limit, $L_{X}/L_{Edd} \simeq 0.001$, we began a new investigation of
the {\it Chandra}/LETGS spectrum of XTE~J1118$+$480 in the low--hard
state (see McClintock et al.\ 2001, Miller et al.\ 2002c).  There is
potential evidence for a cool X-ray disk remaining at the ISCO in this
spectrum also.  However, to make a robust determination, remaining
uncertainties in the instrumental response and spectrum itself must be
investigated in full, and we will report on this analysis in a later
paper.  

\section{Discussion}

\subsection{On the Accretion Flow Geometry in the Low--Hard State}

We have conducted an analysis of {\it XMM-Newton} and {\it RXTE}
spectra of GX~339$-$4 in a low--hard state.  Our results suggest that
a standard cool accretion disk may extend to the innermost stable
circular orbit in the low--hard state.  In a brief analysis of
archival {\it ASCA} spectrum, we find that similar results are readily
obtained for Cygnus X-1.  Moreover, a spectrum recently obtained from
the black hole candidate SWIFT J1753.5$-$0127 (Miller, Homan, \&
Miniutti 2006) during the decline of its outburst appears to support
this disk geometry.  Although we observed GX~339$-$4 in a rising
phase, SWIFT J1753.5$-$0127 was observed during outburst decline, and
Cygnus X-1 is a persistent source, suggesting that disks may commonly
remain at or close to the ISCO in bright phases of the low--hard
state.

These results have a number of consequences.  Transitions between
low--hard and high--soft states in black hole binaries may not
necessarily signal a change in inner disk radius.  Moreover, in
apparent contrast to the absence of jets in disk-dominated high--soft
states (see, e.g., Fender, Belloni, \& Gallo 2004), it may be possible
to maintain a compact, steady jet while the disk remains at the
innermost stable circular orbit.  Homan et al.\ (2001) noted that
state transitions may be partly related to changes in the corona; our
results strengthens this suggestion.

The idea that a transition into the low--hard state marks the onset of
an ADAF with a truncated inner disk (e.g., Esin, McClintock, \&
Narayan 1997) has become a near-paradigm in studies of accreting black
holes.  Our results suggest that an optically-thick accretion disk
with properties closely related to disks observed at higher inferred
accretion rates can operate in the low--hard state.  However, simple
theoretical considerations suggest that it is very unlikely that a
standard thin disk is part of the inner accretion flow in quiescence.
Observational evidence also appears to rule-out the possibility of a
standard accretion disk at the ISCO in quiescence (see, e.g.,
McClintock, Horne, \& Remillard 1995).  By extension, then, our
findings suggest that the accretion flow geometry in quiescence may
not be a simple extrapolation of the geometry of the flow in the
low--hard state.

Fits to the spectra with reflection models reveal a covering fraction
significantly less than unity (see Table 2), consistent with prior
disk reflection fits in the low--hard state (e.g. Gierlinski et al.\
1997).  While a recessed disk would serve to give less disk
reflection, this possibility is inconsistent with our results.  A hard
component which is mildly beamed away from the disk provides a
plausible way to reconcile these findings.

Beloborodov (1999) described a model wherein the corona is fed by
magnetic flares from the disk, and the height of such flares
determines the nature of the corona.  In the low--hard state, the
flares reach mildly relativistic velocities ($v/c \simeq 0.3$),
sufficient to mildly beam the hard X-ray emission away from the disk.
A broadly similar scenario has been described by Merloni \& Fabian
(2002); in that work, the ability of a magnetically--dominated corona
to launch jets is considered.  These models may give insight into how
disks might be connected to jets through a corona, and/or give a sense
of how a corona might act as the base of a jet.  Separately, the
possible role of a jet in producing hard X-ray emission in the
low--hard state has been discussed in detail.  Markoff, Falcke, \&
Fender (2001) suggested that all of the hard X-ray emission in
XTE~J1118$+$480 may be due to the steady, compact jet implied by radio
observations.  Subsequent studies have focused more on the role of
synchrotron self-Comptonization; in this sense, they partially bridge
the gap between synchrotron--dominated jet models and traditional
thermal Comptonization models of the corona.  Markoff \& Nowak (2004)
have showed that the level of disk reflection predicted by
jet-dominated emission models is consistent with observations.

While our deep observation of GX~339$-$4 may provide the most
convincing evidence that a standard thin disk is important in the
low--hard state, it is not the first such evidence.  Prior reports of
cool disks in the low--hard state have sometimes been overlooked
and/or regarded skeptically.  A cool soft excess in the low--hard
state spectrum of Cygnus X-1 was previously reported by Ebisawa et
al.\ (1996), in the same {\it ASCA} spectrum we considered above.  The
presence of a cool thermal component in the low--hard state of Cygnus
X-1 can be traced back much farther: Barr \& van der Woerd (1990) and
Balucinska \& Hasinger (1991) both reported that a soft excess was
required to describe the low--hard state spectrum of Cygnus X-1.  The
interpretation of this component as a disk was complicated by the
possibility that the black hole may partially accrete from the wind of
its massive companion.  However, such a wind is clearly not present in
GX~339$-$4, and the evidence for a disk in the low--hard state of
GX~339$-$4 supports the disk interpetation of the soft component in
Cygnus X-1.

If a disk interpretation of this component was doubted
because the black hole may partially accrete via the wind from its
massive companion, the clear evidence for a disk in the low-mass
binary GX~339$-$4 should make its interpretation clear.

It is likely that current views of the low--hard state have been
heavily influenced by recent observations with {\it RXTE}.  While the
flexibility of {\it RXTE} has made it possible to study the bright
phases of accreting sources with densely-spaced observations, our
results demonstrate that its effective lower energy threshold of 3~keV
makes it insensitive to cool thermal components.  For instance,
Zdziarski et al.\ (2004) address the accretion flow as a function of
state in GX~339$-$4 using {\it RXTE} data; Figure 10 in that work
indicates that the disk suddenly and sharply changes its inner radius
across state transitions.  Such marked changes in apparent inner disk
radius are more likely driven by instrumental limitations, and/or by
modeling assumptions, than physical properties.

If a mechanism such as conduction can radially truncate a disk by
driving rapid evaporation (e.g., Meyer-Hofmeister \& Meyer 1999), it
is clear that it does not necessarily activate with the transition to
the low--hard state.  In the absence of such a mechanism, a
standard thin disk which is truncated interior to 50--100~$R_{g}$ is
not necessarily a stable configuration.  Assuming an accretion
efficiency of 10\% and taking only the mass accretion rate implied by
the disk flux in our observations of GX~339$-$4, the radial inward
drift caused by viscosity in a standard disk, should transfer material
to the innermost stable circular orbit from 50--100~$R_{g}$ in
approximately $10^{3}$~s.  While much longer than the free-fall time
and the orbital timescale at 50--100~$R_{g}$, this drift timescale is
still quite short, and serves to demonstrate how easily a disk can
begin to fill an evacuated inner volume.

In future work on the low--hard state, it will be important to try to
reconcile apparent UV disk components like that found in
XTE~J1118$+$480 (e.g., McClintock et al.\ 2001) with cool X-ray disk
components.  If the two cannot be reconciled, it is possible that soft
X-ray excesses which appear to be cool disks may arise via other means.  It
is also possible that apparent UV disk components are not pure disk
components, and possible that such components are not well understood.
In a survey of UV spectra from short-period black hole binaries, Hynes
(2005) finds that the spectra are generally steeper than the canonical
$\nu^{1/3}$ disk spectrum, and that the broad-band optical spectra are
consistent with power-law forms.  Studies of the IR, optical, and UV
variability observed from XTE~J1118$+$480 in its low--hard state
indicate a synchrotron origin for the variability (Hynes et al.\
2003), suggesting that UV emission in the low--hard state is not
solely due to a disk.

\subsection{Reconciling Timing and Spectral Phenomena}

It is sometimes argued that the low-frequency QPOs (on the order of
1~Hz, and lower) occasionally observed in the low--hard state, serve
to indicate that the disk is radially recessed.  Such arguments are
largely based on the assumption that the low QPO frequency reflects
the Keplerian orbital frequency of the inner disk.  This inference
would certainly contradict spectroscopic evidence that the inner disk
extends to the ISCO, at least in the bright phases of the low--hard
state.  Since the nature of low frequency QPOs is an open question, it
is not possible to entirely refute the possibility that these QPOs
signal a recessed disk.  A few simple considerations, however, argue
against interpreting low-frequency QPOs as evidence for a recessed
disk.

First, the QPOs observed are {\it X-ray} QPOs, and energetic arguments
demand that accreting material release most of its energy in the inner
few Schwarzchild radii.  A disk truncated at several tens or hundreds
of Schwarzschild radii, but which still contributes substantial X-ray
emission, is problematic from the standpoint of producing strong X-ray
oscillations in the {\it RXTE} bandpass.  Embedding the disk in a
Comptonizing corona might solve this problem, and may even explain the
fact that the strength of QPOs is often observed to increase with
energy.  However, in ADAF plus truncated disk models, most
Comptonization occurs interior to the truncated disk, and embedding a
disk in that volume is inconsistent with the basic premise of the
model.  

Second, although high frequency QPOs are apparently absent in the
low--hard state, it is not necessarily appropriate to associate the
Keplerian frequency at the inner disk with low-frequency QPOs.
Low-frequency QPOs similar to those found in low--hard states are
often observed {\it simultaneously} with high-frequency QPOs in the
``very high'' and ``intermediate'' states (also known as ``steep
power-law'' states, or, more generically, bright hard states; see
McClintock \& Remillard 2005).  In such cases, low-frequency QPOs are
{\it not} used to infer the inner disk radius; it is inconsistent to
interpret low-frequency QPOs in this way in the low--hard state.  We
note that it is possible that high frequency QPOs are actually present
in the low/hard state, but have simply not been detected due to a lack
of sensitivity.  We know of no work which has demonstrated that high
frequency QPOs are absent in the low/hard state, at an rms upper limit
similar to the rms at which high-frequency QPOs are detected in other
states.  Far tighter limits will be required to also rule-out the
possibility that the high frequency QPO formation mechanism depends
more strongly on the mass accretion rate than the inner disk radius.

These considerations do not argue against the possibility that QPO
frequency trends (see, e.g., Rossi et al.\ 2004, Belloni et al.\ 2005)
signal a real geometrical change in black hole systems.  It is
possible that changes in the size of the corona, the distance between
the inner disk and jet, the scale height of flares above the disk, or
the timescale for superorbital variations in the disk itself may be
reflected in such frequency trends.  A large number of simultaneous
multi-wavelength observations of black holes in the near future may
help to reveal the nature of low frequency QPOs.

\subsection{On Disk--Jet Connections in Black Holes}

Fender, Belloni, \& Gallo (2004) have recently proposed a broad model
for jet production in black hole binary systems.  This model succeeds
in accounting for many of the observed radio and X-ray properties in
these systems.  Among the most important things any such model needs
to address, is the ubiquitous production of jets in low--hard states,
and the absence of jets in high--soft states, when the disk is
strongest.  Indeed, the absence of jets in the most disk--dominated
states demonstrates that disk--jet coupling may be complex.  

Within the current framework of the Fender, Belloni, \& Gallo (2004)
model, the absence of jet production in the high--soft state is
associated with a smaller inner disk radius in this state.  Our
results indicate that the inner disk radius may not change across
state transitions, and suggest that another parameter is likely
responsible for quenching jets in the high--soft state.  For instance,
if the corona is central to driving jets, the ratio of energy
dissipated in the corona to that dissipated in the disk may be
critical.  A corona which is efficiently cooled below a certain
electron temperature by the disk may be less able to drive jets than a
very hot corona.  Detailed spectroscopy across a black hole outburst
could test whether or not the empirical ``jet line'' found in
hardness--intensity diagrams can be explained in terms of this ratio.
This parameter, and others, may naturally fit within the larger
framework of the Fender, Belloni, \& Gallo (2004) model.

It is also possible that there may be {\it no} simple disk--jet
connection in accreting black holes.  It has already been noted that
if the jet has any role to play in X-ray emission, correlations
between radio and X-ray luminosity may not be disk--jet connections,
but jet--jet connections (Markoff et al.\ 2003).  If the jet does not
play a significant role in X-ray emission, such relations may portray
corona--jet connections (though if the corona is indeed the base of a
jet, whether such relations represent jet--jet connections or
corona--jet connections is largely a question of semantics).  Our
results also suggest that there is no clear relation between hard
X-ray emission and inner disk radius, regardless of the hard X-ray
emission mechanism, so any corona--jet connections could be quite
independent of the inner radius of the disk.

\subsection{Implications for Low--luminosity AGN and LINERs}

It is sometimes suggested that stellar-mass black holes in the
low--hard state, and low-luminosity AGN (LLAGN) and LINERs may have a
common inner accretion flow geometry (e.g., Markowitz \& Uttley 2005).
Although many LLAGN are likely observed at lower Eddington fractions
($10^{-5}$--$10^{-4}$) than the range covered by the three
stellar-mass black hole observations considered in this work, if LLAGN
are truly supermassive black holes in the low--hard state, our results
suggest that the accretion disks in these sources may also remain at
the ISCO.  This suggestion adds to recent evidence that disks are not
dramatically recessed in LLAGN and LINERs.  Maser emission from the
accretion disk in NGC 4258 demands that any inner advective region can
be no larger than 100~$R_{Schw.}$ (Herrnstein et al.\ 1998).
Moreover, a recent UV variability study has detected emission which
can be associated with a disk in a number of LINERs (Maoz et al.\
2005).  Extremely deep X-ray observations of nearby LLAGN such as M81*
and NGC 4258 may be required to determine whether the disks in these
sources are recessed, or if they remain at the ISCO.

\section{Conclusions}

A long {\it XMM-Newton} observation of GX~339$-$4 has demonstrated
that a standard thin accretion disk can remain at the innermost stable
circular orbit around a black hole in the low--hard state.  This
finding has a few important consequences.  Advection-dominated
accretion flows with truncated thin disks may be important at very low
fractions of the Eddington limit, but do not appear to describe the
low--hard state at $L_{X}/L_{Edd} \geq 0.01$.  Similarly, our results
suggest the production of steady compact jets in the low--hard state,
and the quenching of jets in the high--soft state, are phenomena
unrelated to the inner radius of the accretion disk.  Finally, our
spectral results provide additional support for envisioning the hard
X-ray emitting ``corona'' as being related to the base of a jet.


\hspace{0.1in}

We thank Omer Blaes, Michael Garcia, Rob Fender, Sera Markoff, Jeff
McClintock, Ramesh Narayan, Mike Nowak, Chris Reynolds, and Henk
Spruit for stimulating discussions.  We thank Patrick Woudt, Retha
Pretorius, Tetsuya Nagata, and Ikuru Iwata for their contributions to
the multi-wavelength campaign.  We thank the anonymous referee
for constructive comments which improved the paper.  JMM acknowledges
research funding related to this program from NASA.  DS acknowledges
support through the Smithsonian Astrophysical Observatory Clay
Fellowship.  This work is based on observations obtained with {\it
XMM-Newton}, an ESA mission with instruments and contributions
directly funded by ESA member states and the US (NASA).  The Australia
Telescope Compact Array is part of the Australia Telescope which is
funded by the Commonwealth of Australia for operation as a National
Facility managed by CSIRO.  This work has made use of the tools and
services available through HEASARC online service, which is operated
by GSFC for NASA.

\pagebreak

\centerline{~\psfig{file=f1.ps,width=3.2in}~}
\figcaption[h]{\footnotesize A power density spectrum of GX~339$-$4
obtained with {\it RXTE} during the long {\it XMM-Newton} observation
is shown above.  This power spectrum is typical of the low--hard
state, in that it shows high fractional variability and
band-limited noise.  The power spectrum was fit with three Lorentzians
($\chi^{2}/\nu = 204/177$) shown with dashed and dotted lines.}
\medskip

\centerline{~\psfig{file=f2.ps,width=3.2in,angle=-90}~}
\figcaption[h]{\footnotesize A simple absorbed power-law fit to {\it
RXTE} spectra of GX 339$-$4 is shown above.  The 4.0--7.0~keV region
was ignored during the fit.  The {\it RXTE} continuum is clearly
well-described by this simple phenomenological function, with no clear
evidence of a break or cut-off in the spectrum.  Evidence for a broad
Fe~K line can be seen in the data/model ratio.  The lack of obvious
curvature near 30~keV indicates that any disk reflection is likely to
be relatively weak.}  \medskip

\centerline{~\psfig{file=f3.ps,width=3.2in,angle=-90}~}
\figcaption[h]{\footnotesize The ratio of the {\it XMM-Newton} spectra
of GX 339$-$4 when fit with a power-law in the 3-10~keV band are shown
above.  Spectra from revolution 782 are shown in red; spectra from
revolution 783 are shown in blue.  The soft thermal excess can be fit
with an accretion disk which extends to the ISCO, indicating that the
disk is not radially recessed in the low--hard state.  RXTE, with an
effective lower energy bound of 3~keV, has failed to detect such disk
components because the disk contributes minimal flux above 3~keV.}
\medskip

\centerline{~\psfig{file=f4.ps,width=3.2in,angle=-90}~}
\figcaption[h]{\footnotesize {\it XMM-Newton} and {\it RXTE} spectra
of GX~339$-$4 are shown above, fit jointly with a simple absorbed disk
plus power-law model.  {\it XMM-Newton} spectra from revs. 782 and 783
are shown in blue and red, respectively, and the {\it RXTE} spectra
are shown in black.  The 4.0--7.0~keV band was ignored in fitting the
spectra.  The broad Fe~K line is common between the instruments, and is
clearly revealed in the {\it XMM-Newton} spectra. } \medskip

\centerline{~\psfig{file=f5.ps,width=3.2in,angle=-90}~}
\figcaption[h]{\footnotesize The relativistic Fe~K line revealed in
Figure 4 is shown in detail above.  The plot shown above was made in
a manner chosen to prevent biasing the line profile.  The model was a
simple disk plus power-law model, and the 4.0-7.0~keV region was
ignored in fitting the data.  {\it XMM-Newton} data from revs. 782 and
783 are shown in red and blue, respectively, and the {\it RXTE}
spectrum is shown in black.}  \medskip

\centerline{~\psfig{file=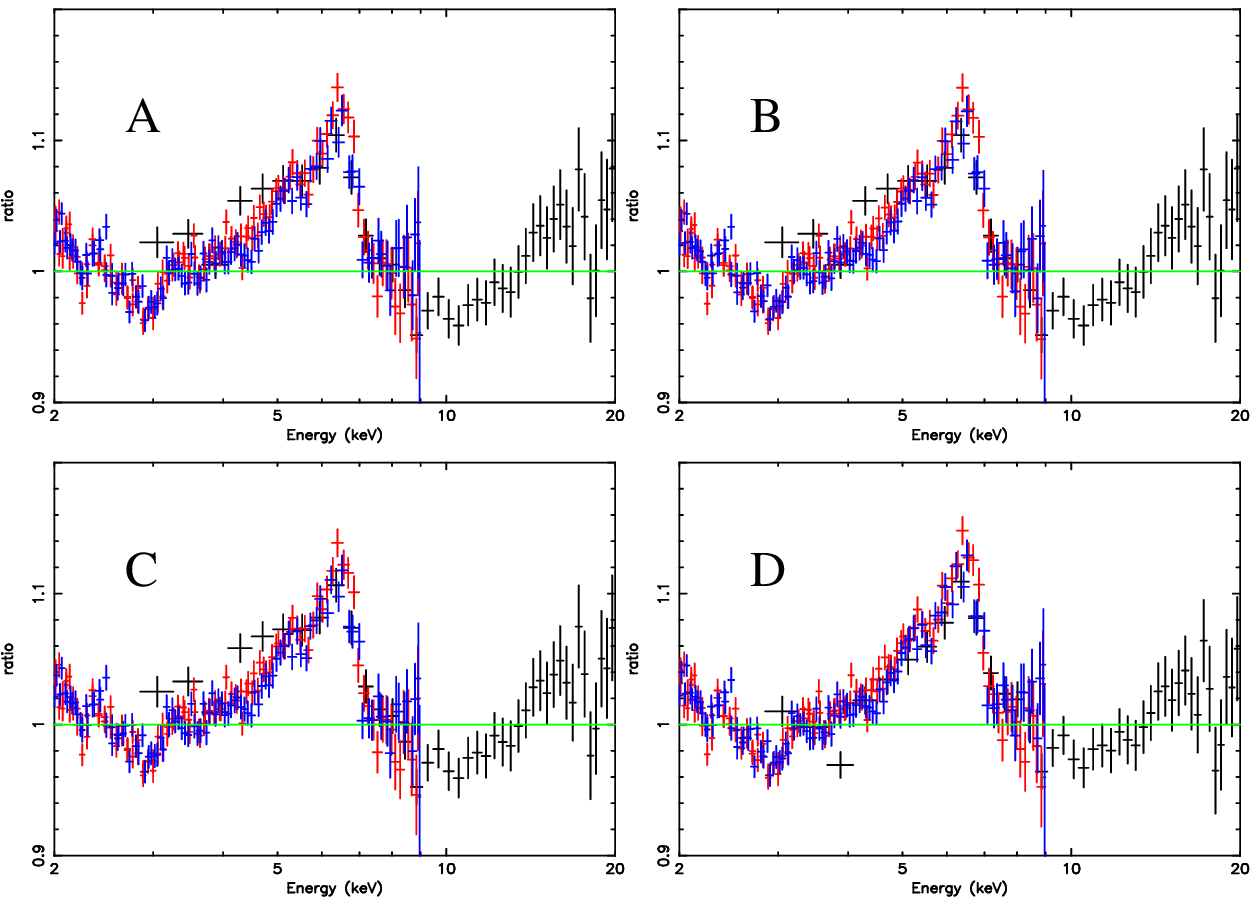,width=3.2in}~}
\figcaption[h]{\footnotesize The relativistic line profile observed in
the low--hard sate of GX~339$-$4 is remarkably independent of the
continuum assumed.  {\it XMM-Newton} spectra from revs. 782 and 783
are shown in red and blue, respectively, and the {\it RXTE} spectra
are shown in black.  In panel A, the the standard ``diskbb'' plus
power-law continuum model was used to form the data/model ratio.  In
panel B, a different disk model, ``diskpn'', was used.  In panel C,
the ``compTT'' Comptonization model was used instead of a power-law.
In panel D, the ``bulk motion Comptonization'' model was used instead
of both additive components.  In all cases, the 4.0--7.0~keV band was
ignored in fitting the spectra.}  \medskip

\centerline{~\psfig{file=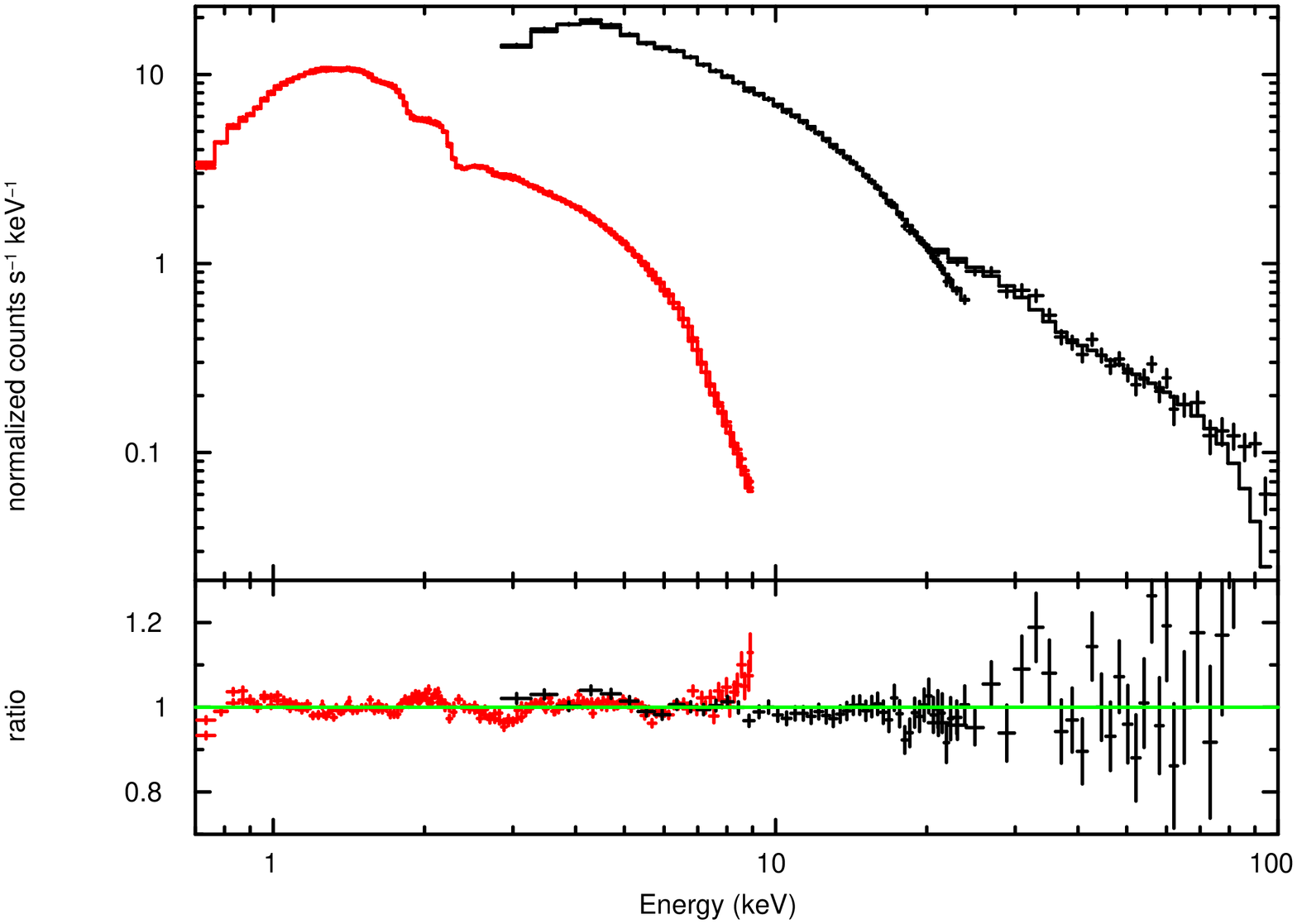,width=3.2in,angle=-90}~}
\figcaption[h]{\footnotesize The plot above shows the result of joint
spectral fits to the simultaneous {\it XMM-Newton} and {\it RXTE}
spectra with the ``constant density ionized disk'' reflection model on
the 0.7--100.0~keV band.  The disk reflection spectrum was
relativistically ``blurred'' with the line function expected from a
disk around a Kerr black hole.  The fit indicates a low reflection
fraction, $f \simeq 0.3$ (where $f \propto \Omega/2\pi$), yet demands
that the disk is within $5~GM/c^{2}$ of the black hole.  This fit
suggests that the hard X-ray component which illuminates the disk may
not be isotropic, but may be mildly beamed away from the disk, as per
the base of a jet.}  \medskip

\centerline{~\psfig{file=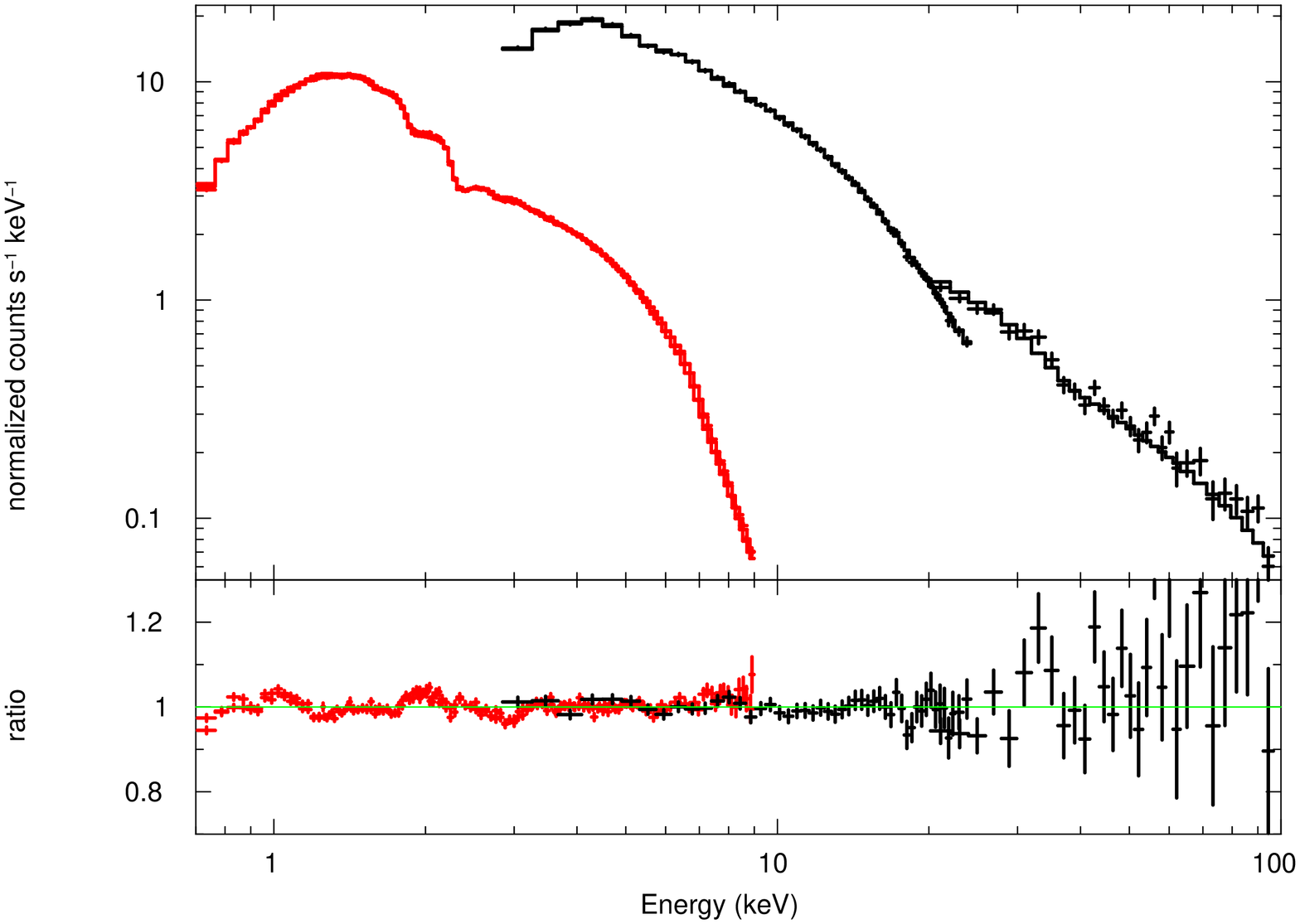,width=3.2in,angle=-90}~}
\figcaption[h]{\footnotesize The plot above shows the result of joint
spectral fits to the simultaneous {\it XMM-Newton} and {\it RXTE}
spectra with the ``pexriv'' disk reflection model.  The results of
this model confirm those obtained with the ``constant density ionized
disk'' model, in that a low reflection fraction and a disk close to
the ISCO is required to describe the data.  This result also suggests
that the ``corona'' may be base of a mildly relativistic jet which
weakly beams the hard X-ray emission.}  \medskip

\centerline{~\psfig{file=f9.ps,width=3.2in,angle=-90}~}
\figcaption[h]{\footnotesize The plot above shows an {\it ASCA}/GIS
spectrum of Cygnus X-1 in the low--hard state, fitted with a simple
power-law in the 3.0--10.0~keV band.  Similar to the {\it XMM-Newton}
spectra of GX~339$-$4, a soft excess is revealed that is consistent
with a cool accretion disk at the ISCO.}  \medskip

\centerline{~\psfig{file=f10.ps,width=3.2in,angle=-90}~}
\figcaption[h]{\footnotesize The plot above shows the ratio of an {\it
ASCA}/GIS spectrum of Cygnus X-1 in the low--hard state, fitted with a
simple absorbed disk blackbody plus power-law model.  The 4.0-7.0~keV
region was ignored in fitting the spectrum.  Similar to the {\it
XMM-Newton} observation of GX~339$-$4, a broad, relativistic Fe~K line
is revealed.  Fits to the line suggest that it originates at radii
consistent with the ISCO, again demanding the disk is not recessed in
the low--hard state of Cygnus X-1.}  \medskip

\clearpage

\begin{table}[t]
\caption{Joint Spectral Fits with Simple Models}
\begin{footnotesize}
\begin{center}
\begin{tabular}{llllllllll}
\tableline
\multicolumn{2}{l}{~} & \multicolumn{2}{l}{diskbb $+$ power-law} & \multicolumn{2}{l}{diskpn $+$ power-law} & \multicolumn{2}{l}{diskbb $+$ compTT} & \multicolumn{2}{l}{bulk. mot. compt.}\\

\tableline

\multicolumn{2}{l}{$N_{H}~(10^{21}~{\rm cm^{-2}})$} & \multicolumn{2}{l}{3.72(5)} & \multicolumn{2}{l}{3.73(4)} & \multicolumn{2}{l}{3.1(1)} & \multicolumn{2}{l}{3.0(1)} \\

\tableline

\multicolumn{2}{l}{$kT_{disk}$~(keV)} & \multicolumn{2}{l}{0.38(1)} & \multicolumn{2}{l}{0.36(2)} & \multicolumn{2}{l}{0.29(1)} & \multicolumn{2}{l}{--} \\
\multicolumn{2}{l}{$R_{disk}~(R_{g})$} & \multicolumn{2}{l}{2.4(2)} & \multicolumn{2}{l}{ 6.0 } & \multicolumn{2}{l}{5(3)} & \multicolumn{2}{l}{ -- } \\
\multicolumn{2}{l}{$N_{disk}$} & \multicolumn{2}{l}{640(80)} & \multicolumn{2}{l}{0.011(2)} & \multicolumn{2}{l}{2700(400)} & \multicolumn{2}{l}{--} \\

\tableline

\multicolumn{2}{l}{$E_{line}$~(keV)} & \multicolumn{2}{l}{6.85(5)} & \multicolumn{2}{l}{6.90(5)} & \multicolumn{2}{l}{6.85(5)} & \multicolumn{2}{l}{6.92(5)} \\
\multicolumn{2}{l}{$q_{line}$} & \multicolumn{2}{l}{3.3(1)} & \multicolumn{2}{l}{3.3(1)} & \multicolumn{2}{l}{3.3(1)} & \multicolumn{2}{l}{3.3(1)} \\
\multicolumn{2}{l}{$R_{line}~(R_{g})$} & \multicolumn{2}{l}{2.8(2)} & \multicolumn{2}{l}{2.8(2)} & \multicolumn{2}{l}{3.1(2)} & \multicolumn{2}{l}{3.1(2)} \\
\multicolumn{2}{l}{$i~({\rm deg})$} & \multicolumn{2}{l}{$18(5)$} & \multicolumn{2}{l}{$18^{+6}_{-8}$} & \multicolumn{2}{l}{$24(8)$} & \multicolumn{2}{l}{$19(9)$} \\
\multicolumn{2}{l}{$N_{line}~(10^{-3})$} & \multicolumn{2}{l}{7.4(5)} & \multicolumn{2}{l}{7.3(5)} & \multicolumn{2}{l}{7.6(5)} & \multicolumn{2}{l}{8.0(7)} \\
\multicolumn{2}{l}{$W_{line}$~(eV)} & \multicolumn{2}{l}{350(20)} & \multicolumn{2}{l}{350(20)} & \multicolumn{2}{l}{360(20)} & \multicolumn{2}{l}{380(40)} \\

\tableline

\multicolumn{2}{l}{$\Gamma$} & \multicolumn{2}{l}{1.46(1)} & \multicolumn{2}{l}{1.46(1)} & \multicolumn{2}{l}{--} & \multicolumn{2}{l}{--} \\
\multicolumn{2}{l}{$kT_{0}$} & \multicolumn{2}{l}{--} & \multicolumn{2}{l}{--} & \multicolumn{2}{l}{0.27(1)} & \multicolumn{2}{l}{0.26(1)} \\
\multicolumn{2}{l}{$kT_{e}$} & \multicolumn{2}{l}{--} & \multicolumn{2}{l}{--} & \multicolumn{2}{l}{21(2)} & \multicolumn{2}{l}{--} \\
\multicolumn{2}{l}{$\tau$} & \multicolumn{2}{l}{--} & \multicolumn{2}{l}{--} & \multicolumn{2}{l}{2.8(1)} & \multicolumn{2}{l}{--} \\
\multicolumn{2}{l}{$\alpha$} & \multicolumn{2}{l}{--} & \multicolumn{2}{l}{--} & \multicolumn{2}{l}{--} & \multicolumn{2}{l}{0.48(1)} \\
\multicolumn{2}{l}{log(A)} & \multicolumn{2}{l}{--} & \multicolumn{2}{l}{--} & \multicolumn{2}{l}{--} & \multicolumn{2}{l}{0.80(2)} \\
\multicolumn{2}{l}{$N_{hard}$} & \multicolumn{2}{l}{0.320(5)} & \multicolumn{2}{l}{0.320(5)} & \multicolumn{2}{l}{0.042(3)} & \multicolumn{2}{l}{0.0150(2)} \\

\tableline

\multicolumn{2}{l}{$\chi^{2}/\nu$} & \multicolumn{2}{l}{3899.2/2256} & \multicolumn{2}{l}{3905.9/2256 } & \multicolumn{2}{l}{4071.0/2255 } & \multicolumn{2}{l}{4103.1/2259} \\

\tableline
\end{tabular}
\vspace*{\baselineskip}~\\ \end{center} 
\tablecomments{The table above details the results of fitting simple
spectral models to all four {\it XMM-Newton}/EPIC-MOS and two {\it
RXTE} PCA and HEXTE spectra jointly with simple continuum models.  The
``diskbb'' model is the standard multicolor disk model, and the
``diskpn'' model is a variation which assumes a pseudo-Newtonian inner
potential (see Zimmerman et al.\ 2005 for a discussion).  The ``Laor''
disk line component was used in the above models.  Inner disk radii
inferred via the disk continuum models assume $i = 18^{\circ}$ and a
distance of 8.5~kpc, and are subject to various systematic errors (see
the text).  Unlike the other three models, the ``bulk motion
Comptonization'' model is a single continuum, and does not contain a
separated disk component.  Symmetric errors are given in parentheses;
single-digit errors indicate the error in the last significant digit
for the given parameter value.}
\vspace{-1.0\baselineskip}
\end{footnotesize}
\end{table}

\begin{table}[t]
\caption{Joint Spectral Fits with Relativistically--blurred Disk Reflection Models}
\begin{footnotesize}
\begin{center}
\begin{tabular}{llllll}
\tableline
\multicolumn{2}{l}{~} & \multicolumn{2}{l}{const. dens. ion. disk} & \multicolumn{2}{l}{perxiv}\\

\tableline

\multicolumn{2}{l}{$N_{H}~(10^{21}~{\rm cm^{-2}})$} & \multicolumn{2}{l}{4.4(4)} & \multicolumn{2}{l}{3.7(4)} \\

\tableline

\multicolumn{2}{l}{$kT_{disk}$~(keV)} & \multicolumn{2}{l}{0.30(4)} & \multicolumn{2}{l}{0.39(4)} \\
\multicolumn{2}{l}{$R_{disk}~(R_{g})$} & \multicolumn{2}{l}{ ~ } & \multicolumn{2}{l}{ ~ } \\
\multicolumn{2}{l}{$N_{disk}$} & \multicolumn{2}{l}{2100(200)} & \multicolumn{2}{l}{700(200)} \\

\tableline

\multicolumn{2}{l}{$E_{line}$~(keV)} & \multicolumn{2}{l}{--} & \multicolumn{2}{l}{6.8(1)} \\
\multicolumn{2}{l}{$q_{line/blur}$} & \multicolumn{2}{l}{3.0} & \multicolumn{2}{l}{3.0} \\
\multicolumn{2}{l}{$R_{line/blur}~(R_{g})$} & \multicolumn{2}{l}{5.0(5)} & \multicolumn{2}{l}{4.0(5)} \\
\multicolumn{2}{l}{$i~({\rm deg.})$} & \multicolumn{2}{l}{$20^{+5}_{-10}$} & \multicolumn{2}{l}{$20^{+5}_{-15}$} \\
\multicolumn{2}{l}{$N_{line}~(10^{-3})$} & \multicolumn{2}{l}{--} & \multicolumn{2}{l}{3.5(3)} \\

\tableline

\multicolumn{2}{l}{$\Gamma$} & \multicolumn{2}{l}{1.50} & \multicolumn{2}{l}{1.41(3)} \\
\multicolumn{2}{l}{$f \propto \Omega/2\pi$} & \multicolumn{2}{l}{0.31(3)} & \multicolumn{2}{l}{0.22(6)} \\
\multicolumn{2}{l}{log($\xi$)} & \multicolumn{2}{l}{2.8(1)} & \multicolumn{2}{l}{3.0} \\
\multicolumn{2}{l}{$N_{hard}$} & \multicolumn{2}{l}{$1.9(3)\times 10^{-25}$} & \multicolumn{2}{l}{0.32(3)} \\

\tableline

\multicolumn{2}{l}{$\chi^{2}/\nu$} & \multicolumn{2}{l}{2055.1/1165} & \multicolumn{2}{l}{2120.5/1160} \\

\tableline
\end{tabular}
\vspace*{\baselineskip}~\\ \end{center} 
\tablecomments{The table above details the results of jointly fitting
relativistically-blurred disk reflection models to the two {\it
XMM-Newton}/EPIC-MOS spectra from revolution 782 and two {\it RXTE}
PCA and HEXTE spectra.  Symmetric errors are given in parentheses.
Single-digit errors indicate the error in the last significant digit
for the given parameter value.  The constant density ionized
reflection model includes line emission, and a separate line component
was not included in the overall spectral model.  The pexriv model does
not include a line, and its blurring parameters were tied to those of a
``Laor'' line included in the overall spectral model.  Where errors
are not given, the parameter was fixed at the quoted value.}
\vspace{-1.0\baselineskip}
\end{footnotesize}
\end{table}

\end{document}